\nonstopmode

\newcommand{\be}{\begin{equation}}
\newcommand{\ee}{\end{equation}}
\newcommand{\sgn}{\mathrm{sgn}}

\documentclass[apl,showpacs,twocolumn]{revtex4}
\usepackage{graphicx}
\usepackage{color}
\usepackage{amsmath}
\usepackage{accents}

\begin{document}
\title{Thermally-Assisted Spin-Transfer Torque Magnetization Reversal of Uniaxial Nanomagnets in Energy Space}
\author{D.~Pinna}
\email{daniele.pinna@nyu.edu}
\affiliation{Department of Physics, New York University, New York, NY 10003, USA}
\author{D.~L.~Stein}
\affiliation{Department of Physics, New York University, New York, NY 10003, USA}
\affiliation{Courant Institute of Mathematical Sciences, New York University, New
             York, NY 10012, USA}
\author{A.~D.~Kent}
\affiliation{Department of Physics, New York University, New York, NY 10003, USA}

\begin{abstract}

We propose a new analytical method for studying the asymptotic behavior of switching time as a function of current in macrospins under the effects of both spin-torque and thermal noise by focusing on their diffusive energy space dynamics. We test our method on the well understood uniaxial macrospin model and confirm the switching scaling dependence ($I\rightarrow 0$, $\langle\tau\rangle\propto\exp(-\xi(1-I)^2)$). The analysis also shows that when there is an angle between spin current and magnet’s uniaxial axes the mean switching time in the low current limit depends on the spin-current projection on the easy axis, but otherwise has the same functional form as that in the collinear case. These results have implications for modeling the energetics of thermally assisted magnetization reversal of spin transfer magnetic random access memory bit cells.

\end{abstract}

\maketitle

A spin-polarized current passing through a small magnetic conductor is known to deposit its spin-angular momentum, leading to precession and, in some cases, reversal phenomena in the magnetic system~\cite{Slon,Berger}. A thorough understanding of these effects, however, requires taking into account the effect of thermal fluctuations inducing diffusion in the moment of a magnetic material. This is of particular experimental relevance since spin-transfer effects on nanomagnets are often studied at low currents, where noise is expected to dominate. The standard theoretical approach used to study the dynamics of small magnetic structures has been 
to treat the magnetic system as a macrospin in the spirit of Brown~\cite{Brown}, thus modeling the dynamics as those of a magnetic moment undergoing damped precession in an effective field. The inclusion of spin-tranfer torque, after Slonczewski~\cite{Slon}, is included phenomenologically by modifying the macrospin's Landau-Lifshitz-Gilbert (LLG) dynamical equation with the addition of an extra term ensuring that the magnetic body absorbs spin-angular momentum perpendicularly to its orientation. 

The LLG equation for a normalized magnetization $\mathbf{m}$ then reads:

\begin{eqnarray}
\mathbf{\dot{m}}&=&-\gamma'\mathbf{m}\times\mathbf{H}_\mathrm{eff}-\alpha\gamma'\mathbf{m}\times\left(\mathbf{m}\times\mathbf{H}_\mathrm{eff}\right)\nonumber\\
&-&\gamma' j\mathbf{m}\times\left(\mathbf{m}\times\mathbf{\hat{n}}_p\right)+\gamma'\alpha j\mathbf{m}\times\mathbf{\hat{n}}_p,
\end{eqnarray}
where $\gamma'=\gamma/(1+\alpha^2)$, $\gamma$ is the gyromagnetic ratio, $\alpha$ the Landau damping coefficient and $j=(\hbar/2e)\eta J$ is the spin-angular momentum deposited per unit time with $\eta = (J_{\uparrow}-J_{\downarrow})/(J_{\uparrow}+J_{\downarrow})$, the spin-polarization factor of incident current $J$. The last two terms describe a vector torque generated by current polarized in the direction $\mathbf{\hat{n}}_p$. The effective field $\mathbf{H}_\mathrm{eff}$ is immediately obtained once the energy landscape of the magnetic sample is specified. In the simplified case of a uniaxial single domain magnet, it is characterized by the projection of the magnetization onto the uniaxial anistropy axis:

\begin{equation}
U(\mathbf{m})=-K(\mathbf{m}\cdot\mathbf{\hat{n}}_K)^2 - \mathbf{m}\cdot\mathbf{H},
\end{equation}
with $K=(1/2)M_SH_KV$ the uniaxial anisotropy of a magnet of volume $V$ and anisotropy field $H_K$, saturation magnetization $M_S$, applied field $\mathbf{H}$, and $\mathbf{\hat{n}}_K$ the unit vector representing the orientation of the uniaxial anisotropy (easy) axis.

Starting from (1), thermal fluctuations are introduced by considering the effect of a random applied magnetic field with mean zero and variance set by the fluctuation dissipation theorem. Doing this leads to a set of three coupled stochastic LLG equations with multiplicative noise~\cite{Brown,Palacios}. Experimentally, one is generally interested in the magnetization reversal properties of a macrospin in a regime dominated by thermal-activation related processes~\cite{Brataas, Bedau}. Within the macrospin approximation, the relevant thermally assisted spin transfer torque reversal energetics are not well defined as a consequence of the inherently non-conservative nature of the added spin-torque term which does not allow the construction of a generalized energy landscape. 

In the simplified uniaxial macrospin model where the anisotropy and spin polarization axes are chosen to be collinear, and demagnetization effects are neglected, the stochastic evolution equation determining the switching properties of the macrospin can be described by a single 1D uncoupled stochastic differential equation~\cite{Wang1,Taniguchi,APL}. 

In this paper we present a new analysis which, unders suitable assumptions, allows one to reduce the complexity of the LLG dynamics to an equivalent 1D uncoupled equation and to quantify its thermally acitvated behavior succinctly. We believe this to be very useful toward the characterization of switching properties in more complex macrospin models where symmetry arguments cannot be found to simplify the general coupled structure of the magnetization evolution equations. 

To elucidate the most basic properties of thermally-assisted magnetization reversal under spin-transfer torque, we choose to consider a magnetic system solely characterized by its uniaxial anisotropy~\cite{Sun} and focus on its thermally assisted dynamics in the absence of any applied fields. The energy landscape is given by the first term on the right hand side of (2). The equilibrium states in this landscape are then given by parallel and anti-parallel configurations of the magnetization along the uniaxial anisotropy axis.  We consider our magnetic sample to be initially magnetized in thermal equilibrium in one of its two possible stable states (i.e. $m=+\mathbf{\hat{n}}_K$ or $m=-\mathbf{\hat{n}}_K$). Upon applying a (positive) current, spin-torque effects will attempt to drive the magnetization toward aligning with the polarization axis $\mathbf{\hat{n}}_P$. If the current is then switched off again, the magnetization will relax back toward the easy axis. To capture this behavior, it is very helpful to write down the dynamical equation for the projection $q\equiv\mathbf{m}\cdot\mathbf{\hat{n}}_K$ of the magnetization along the easy axis. For temperatures such that $\xi\equiv K/kT\gg1$, the sign of $q$ will specify to a high degree of accuracy to which state the magnetization will relax after the current is turned off. These ``projectional dynamics'' read:

\begin{eqnarray}
\dot{q}&=&\alpha\left[(n_zI+q)(1-q^2)+n_xIqs\right]\nonumber\\
&+&\alpha^2In_xm_y+\sqrt{\frac{\alpha}{\xi}(1-q^2)}\circ\dot{W},
\end{eqnarray}
where $I=j/(\alpha H_K)$ is the normalized current and a natural time unit $\tau=\gamma'H_K t$ has been introduced. If we choose $\mathbf{\hat{n}}_P$ as our z-axis, and $\mathbf{\hat{n}}_K$ oriented somewhere in the x-z plane, then $s$ is the projection of the dynamics on the axis orthogonal to both the y-axis and $\mathbf{\hat{n}}_K$. To simplify the notation, then, $n_x$ and $n_z$ are the components of $\mathbf{\hat{n}}_K$ along the original $\hat{x}$ and $\hat{z}$ axes. The last term appearing in the expression models the effects of thermal noise whose stochastic contribution is intended according to Stratonovich calculus (indicated by the symbol $\circ$) and $\dot{W}$ is a standard mean $0$, variance $1$ Wiener process~\cite{Palacios, Karatsas}.  

In the simplified collinear case, where polarizer and uniaxial axes are aligned ($n_z=1$,$n_x=0$), the projectional dynamics are the same as the regular dynamics for the $m_z$ component of the magnetization, and are decoupled from both $m_x$ and $m_y$. The problem is effectively reduced to that of a 1D stochastic differential equation characterizing the switching behavior over all possible currents~\cite{APL}.

As already hinted, unless simplifying symmetries are invoked, the main complication in studying a macrospin's dynamical behavior lies in the generally complex structure of its coupled set of stochastic differential equations. Equation (3) is precisely a realization of such: in the case of collinearity between uniaxial and polarizer axes, the evolution equation for $m_z\equiv q$ does not depend on any of the other two magnetization components. The moment any such symmetry is broken, or more complex energy landscapes are considered (such as considering biaxial anisotropy scenarios), the full set of coupled equations (see (3) in case of general, non-zero tilt) must be considered. 

The evolution of a nanomagnet is expected to operate in a regime where the nonconservative dynamics act on timescales that are much longer than that of the energy-conserving part of the hamiltonian. This can be established by noting first that currents $I$ governing the criticality of (3) are $O(1)$ and, second, all terms with the exception of the first in (1) are of order $\alpha$ or higher. As such, for $\alpha\ll 1$, the two evolution timescales are separated. Trajectories then should remain close to stable periodic Hamiltonian orbits, and slowly drift from one orbit to another due to damping, STT and thermal noise. In ~\cite{APL}, the authors addressed the problem of broken symmetries due to the presence of tilt by averaging equation (3) over constant energy trajectories. In doing so, they again retained a 1D stochastic differential equation and solved its associated mean first passage time problem. This approach, though, is just a particular example (see also~\cite{Wang2,TaniguchiE}) of a more general analysis which can be applied to macrospin dynamics. 

Take as an example the form of the energy landscape in (2) in the absence of applied magnetic fields and ask how it evolves in time. Taking a time derivative and using the definition of $q$, one finds:

\be
\dot{\epsilon}\equiv\frac{\dot{U}}{K}=-2q\dot{q}.
\ee
Combining (3) and (4) and averaging the dynamics over constant energy trajectories leads to a diffusion equation in energy space~\cite{Apalkov,Katie,Serpico}. Such an equation is one dimensional and can be defined independently of the complexity of the energy landscape for any generalized macrospin system. In our particular case, the averaging is very straightforward to accomplish since constant energy trajectories ($\alpha\to0$,$I\to0$,$\xi\to\infty$) are simple circular librations around the uniaxial anisotropy axis with $q=\sgn(q)\sqrt{|\epsilon|}$. Proceeding with the averaging and noting that $\langle q^2s\rangle=\langle qm_y\rangle=0$ due to the geometrical structure of the constant energy orbit, we find for states starting antiparallel to the uniaxial anisotropy axis:

\be
\dot{|\epsilon|}=2\alpha\sqrt{|\epsilon|}(1-|\epsilon|)(\sqrt{|\epsilon|}-In_z)+2\sqrt{\frac{\alpha}{\xi}|\epsilon|(1-|\epsilon|)}\circ\dot{W}.
\ee

Two fundamental properties of this energy space evolution are immediately seen. The first is that $In_z=1$ behaves as a crossover value above which all dynamics prefer evolving towards the parallel state independently of noise. Accordingly, this value of $In_z$ is hereafter denoted as the ``critical current''. Analogously, the thermal regime is characterized by sub-critical current values ($In_z\ll 1$). The second is that the angular tilt between uniaxial and polarizer axes factors in only through the term $n_z$ multiplying the current $I$. As such, as the angular tilt is varied, the dynamics are going to differ only up to a rescaling of the current $I$ and, beyond that, all other properties of the dynamics will be functionally identical. This was one of the main results shown in~\cite{APL} where the authors simplified (3) via a method analogous to averaging along constant energy trajectories.  The regime of validity of this procedure is ascertained by imposing that the drift in energy due to noise and spin-torque happen on a timescale much smaller than the precessionary timescale. Imposing then that $\underaccent{\epsilon}{\mathrm{max}}T(\epsilon)|\dot{\epsilon}|\ll1\nonumber$, one obtains as a condition: $I\ll(2\pi\alpha)^{-1}.$

The thermally assisted properties are then easily understood by studying the mean first passage time over the effective barrier to the dynamics of $|\epsilon|$ analogously to what has been done in the literature. Instead, we consider another method (initially due to Friedlin and Wentzell~\cite{FW}) to extract the exponential scaling behavior of the mean first passage time without resorting to the full mean first passage time problem. Formally, given a Langevin equation $\dot{x}=f(x)+g(x)\circ\dot{W}$, the Wiener process $\dot{W}$ can be solved at each instant in time. At each such instant, the probability of a Gaussian-distributed random kick can be written as:

\be
P[\dot{W}]|_t=\exp{\left[-\frac{1}{2}|\dot{W}|^2\right]}=\exp{\left[-\frac{1}{2}\left(\frac{\dot{x}-f(x)}{g(x)}\right)^2\right]}.\nonumber
\ee 
The probability of specific realization of a stochastic path will then simply the product of the probabities of all the random kicks that the path witnessed. As such the probability of a path starting in some point $A$ and ending in $B$ in time $T$ will be:

\be
P[A\to B]=\exp{\left[-\int_0^T dt\frac{1}{2}\left(\frac{\dot{x}-f(x)}{g(x)}\right)^2\right]}.\nonumber
\ee 
The most likely trajectory starting in $A$ and ending in $B$ in time $T$ will be that which maximizes its probability of realization. One then seeks to minimize the factor appearing in the exponential, the F-W action:

\be
S_{\mathrm{FW}}=\int_0^T dt\frac{1}{2}\left(\frac{\dot{x}-f(x)}{g(x)}\right)^2.
\ee
Finding the action of such a trajectory is then a standard Hamilton-Jacobi problem (see Appendix):

\be
\Delta S_{\mathrm{FW}}=S_{\mathrm{FW}}(B)-S_{\mathrm{FW}}(A)=-2\int_A^B\frac{f(x)}{g^2(x)}dx.
\ee

We are interested in the F-W action of the most probable trajectory starting in $|\epsilon|=1$ and ending at the saddlepoint of the deterministic dynamics: $|\epsilon|=I^2$ (we specialize to the case $n_z=1$ since all other cases are immediately reobtained be rescaling). Plugging (5) into (7), one finds that:

\be
\Delta S_{\mathrm{FW}}=\xi\int_{I^2}^1\frac{\sqrt{|\epsilon|}-I}{\sqrt{|\epsilon}|}d|\epsilon|=\xi(1-I)^2.\nonumber
\ee
The exponential scaling between mean switching time and current is then $\tau\propto\exp(\xi(1-I)^2)$~\cite{Wang1,Taniguchi,APL}.

We have proposed a novel analytical approach to assessing the thermally activated behavior of a macrospin system. This is in agreement with work done on uniaxial macrospins in the presence of spin transfer torque. Clearly, the correct asymptotic scaling form is essential to properly determine the energy barrier to reversal. The energy barrier, in turn, is very important in assessing the thermal stability of magnetic states of thin film elements that are being developed for long term data storage in STT-MRAM. Further work should address how these results extend to systems with easy plane anisotropy and situations in which the nanomagnet has internal degrees of freedom, leading to a break down of the macrospin approximation. The method discussed allows to treat analytically even more complex dynamics where other anisotropy contributions wish to be taken into account. To keep the exposition as simple as possible, we have refrained from solving the full mean first passage time problem associated with equation (5). Doing so is straightforward and proceeds exactly along the lines of what was shown in ~\cite{Taniguchi,APL}. The one advantage of solving the full first passage time is that one is capable of extracting information on the exponential prefactor for the reversal time scalings. We will address the application of this methodology to the full biaxial anisotropy model in the presence of spin-torque and thermal noise in a future publication.

\begin{acknowledgments}
The authors would like to acknowledge A. MacFadyen and J. Z. Sun for comments leading to this paper. This research was supported by NSF-DMR-100657 and PHY0965015.

\end{acknowledgments}

\appendix
\section{}

Starting from (6), we wish to calculate the minimum action for a system governed by the effective lagrangian:

\be
L_{FW}=\frac{1}{2}\left(\frac{\dot{x}-f(x)}{g(x)}\right)^2.
\ee

By doing a legendre transformation we can go to the associated hamiltonian form:

\be
H_{FW}(p,x)=p\dot{x}-L_{FW}=p\left(p\frac{g^2(x)}{2}+f(x)\right).
\ee

Since the lagrangian does not depend explicitely on time, the Hamilton-Jacobi equation for the problem reduces to the simple form $H_{FW}(\partial_xS_{FW},x)=0$, where $S_{FW}$ is the action for the problem. This leads to the differential equation for $S_{FW}$:

\be
\partial_xS_{FW}=-2\frac{f(x)}{g^2(x)},
\ee
shown in equation (7) of the text.

\begin{thebibliography}{99}

\bibitem{Slon} J. Slonczewski, J. Magn. Magn. Mater. {\bf159}, L1 (1996).
\bibitem{Berger} L. Berger, Physical Review B {\bf 54}, 9353 (1996).
\bibitem{Brown} W. F. Brown, Phys. Rev. {\bf135}, 5 (1963).
\bibitem{Palacios} J. L. Garcia-Palacios and F. J. Lazaro, Phys. Rev. B {\bf58}, 22 (1998).
\bibitem{Brataas} A. Brataas, A. D. Kent, H. Ohno, Nature Materials {\bf 11}, 372 (2012).
\bibitem{Bedau} D. Bedau {\sl et al.}, Appl. Phys. Lett. {\bf97}, 262502 (2010).
\bibitem{Wang1} X. Wang, W. Zhu, and D. Dimitrov, Phys. Rev. B {\bf78}, 024417 (2008).
\bibitem{Taniguchi} T. Taniguchi and H. Imamura,Phys. Rev. B {\bf83}, 054432 (2011).
\bibitem{APL} D. Pinna, Aditi Mitra, D. L. Stein, and A. D. Kent, Appl. Phys. Lett. {\bf101}, 262401 (2012).
\bibitem{Sun} J. Z. Sun, Phys. Rev. B {\bf62}, 1 (2000).
\bibitem{Karatsas} I. Karatsas and S. Shreve, {\em Brownian Motion and Stochastic Calculus, 2nd ed.}(Springer-Verlag, New York, 1997).
\bibitem{Wang2} X. Wang, W. Zhu, Z. Gao, H. Xi, and D. Dimitrov, J. Appl. Phys. {\bf105}, 07D103 (2009).
\bibitem{TaniguchiE} T. Taniguchi, Y. Utsumi, M. Marthaler, D. S. Golubev, H. Imamura, arXiv:1211.5818 (2012) (preprint) 
\bibitem{Apalkov} D. M. Apalkov and P. B. Visscher, Phys. Rev. B {\bf72}, 180405R (2005).
\bibitem{Katie} K. Newhall and E. Vanden-Eijnden, arXiv:1210.6253 (2012) (preprint)
\bibitem{Serpico} G. Bertotti, I. Mayergoyz, and C. Serpico, {\em Nonlinear Magnetization Dynamics in Nanosystems} (Elsevier, Oxford, UK, 2009).
\bibitem{FW} M. I. Freidlin and A. D. Wentzell {\em Random Perturbations of Dynamical Systems} (Springer, New York, 1991)


\end{thebibliography}
\end{document}